# Correlation Effects in Wave Function Mapping of Molecular Beam Epitaxy Grown Quantum Dots

*Giuseppe Maruccio,[1,*] Martin Janson,[1] Andreas Schramm,[1] Christian Meyer,[1] Tomohiro Matsui,[1] Christian Heyn,[1] Wolfgang Hansen,[1] Roland Wiesendanger,[1] Massimo Rontani,[2,†] and Elisa Molinari[2]*

[1]Institute of Applied Physics, University of Hamburg, Jungiusstrasse 11, 20355 Hamburg, Germany

[2]CNR-INFM National Research Center S3 and Dipartimento di Fisica, Università degli Studi di Modena e Reggio Emilia, Via Campi 213/A, 41100 Modena, Italy

[*]Present address: CNR-INFM National Research Center NNL, Distretto Tecnologico ISUFI Via Arnesano, 73100 Lecce, Italy. Email address: giuseppe.maruccio@unile.it
[†]Email address: rontani@unimore.it

## Abstract

We investigate correlation effects in the regime of a few electrons in uncapped InAs quantum dots by tunneling spectroscopy and wave function (WF) mapping at high tunneling currents where electron-electron interactions become relevant. Four clearly resolved states are found, whose approximate symmetries are roughly *s* and *p*, in order of increasing energy. Because the major axes of the *p*-like states coincide, the WF sequence is inconsistent with the imaging of independent-electron orbitals. The results are explained in terms of many-body tunneling theory, by comparing measured maps with those calculated by taking correlation effects into account.

Quantum dots (QDs) have attracted great attention in the last years as ideal materials where three-dimensional (3D) electronic confinement leads to novel phenomena and applications[1], from optoelectronics to implementation of quantum computation. In particular, strain-induced InAs QDs have been largely investigated and used in optoelectronics due to the possibility of achieving emission at wavelengths of interest for telecommunications[2-5]. The wave functions (WFs) of electrons and holes confined in the QDs are the most basic features ultimately determining all QD properties. Some of us previously demonstrated that it is possible to map the dot WFs in the one-electron regime by means of spatially resolved tunneling spectroscopy images[6]. The presence of more electrons in a QD leads to a Coulomb interaction between carriers. As a consequence the injection of an additional electron into the QD affects its energy spectrum and ultimately changes the WFs, leading to novel ground and excited states. Understanding this basic but fundamental issue would be a key step forward for the comprehension of few-particle interactions in strongly correlated systems as well as for applications in the fields of single-electron devices, spintronics, and quantum information encoding. Indeed, QDs with a few electrons can be strongly interacting objects and large correlation effects have been recently reported in light scattering[7] as well as in high source-drain voltage spectroscopies[8].

Despite theoretical predictions that WF mapping of QDs should be sensitive to correlation effects[9, 10], there has been no clear experimental evidence yet. In fact, all WF images obtained so far, in both real[6, 11, 12] and reciprocal space[13-16], have been basically interpreted in terms of independent-electron orbitals. Here we investigate correlation effects by WF mapping. Specifically, we study uncapped InAs QDs by tunneling spectroscopy at high tunneling currents where electron-electron interaction and correlation effects become relevant. We find four clearly resolved states: their maps loosely display *s*-like and *p*-like symmetries, in order of increasing voltage. Surprisingly, the *p*-like states elongate in the same direction. As a consequence, it is not possible to explain the WF sequence in terms of either one-electron orbitals (then two orthogonal *p*-like states should be observed) or self-consistent orbitals (a replica of the *s*-like instead of the *p*-like state would be expected, in the case of charging of a second electron). We show that correlation effects play a crucial role in this case and we use many-body tunneling theory combined with full configuration interaction (FCI) calculations[17] to explain the experimental observations.

The strain-induced InAs QDs were grown on n-doped GaAs(001) substrates by molecular beam epitaxy (MBE). First, an n-doped GaAs buffer layer ($N_D \approx 2 \times 10^{18}$ cm$^{-3}$) 200 nm thick was deposited at a temperature of 620 °C. Then, an undoped tunneling barrier 5 nm thick ($N_A < 10^{15}$ cm$^{-3}$) was overgrown (see Figure 1a) in order to provide the decoupling necessary to investigate the inherent electronic properties of the QDs and to obtain direct images of the dot WFs disentangled from the electronic structure of the substrate. Finally, the sample was cooled down to 500 °C in order to form the QDs by depositing 2.1 ML of InAs at a growth rate of 0.05 ML/s. Reflection high-energy electron diffraction (RHEED) was used to monitor *in situ* QD formation. A transition from a streaky to a spotty pattern (indicating the onset of three-dimensional islanding) and chevron-like spots generally attributed to QD facets were observed[18]. The base pressures of the MBE and scanning tunneling microscopy (STM) chambers were below 10$^{-10}$ mbar and the samples were transferred into the STM (within 1 h) without being exposed to air by means of a mobile ultrahigh vacuum transfer system at $p < 10^{-9}$ mbar in order to avoid contamination.

To spatially map the energy-resolved local electron density, we have used scanning tunneling spectroscopy (STS). According to the standard mean-field theory[19], the differential tunneling conductivity $dI/dV(V, x, y)$ is to a good approximation proportional to the local density of independent-electron states (LDOS), and in the case of a system described by discrete states $\psi_i(E_i, x, y)$ the LDOS is given by $\sum_{\delta E} |\psi_i(E_i, x, y)|^2$. As a consequence, if the energy resolution $\delta E$ is less than the energy level spacing, the LDOS reduces to a single term and spatially-resolved $dI/dV$ maps display the detailed spatial structure of $|\psi_i(E_i, x, y)|^2$ at the corresponding energy $eV$. If correlation effects are strong, however, the concept of LDOS is not appropriate and $dI/dV(V, x, y)$ is instead expected to be proportional to $-1/\hbar\pi \, \mathrm{Im}\, G(x, y; x, y; eV)$, where $G$ is the interacting retarded Green's function (or single-electron propagator) resolved in both energy and space[9, 10, 20]. The imaginary part of $-G/\hbar\pi$ (known as spectral density) may be regarded as the modulus squared of a *quasi particle* WF $\varphi_i(E_i, x, y)$, which can considerably deviate from its independent-particle counterpart $\psi_i(E_i, x, y)$ (refs 9 and 10).

Our STM operates at $p \ll 10^{-10}$ mbar and $T = 6$ K with a maximum energy resolution of $\delta E = 2$ meV (ref 21). STM images were taken in constant-current mode using W and PtIr tips with a typical sample bias in the range of 2-4 V and a tunneling current of 20-40 pA. A lock-in technique (modulation voltage $V_{mod}$ in the range of 5-20 mV) was used to record $dI/dV(V, x, y)$ and WF mapping was carried out over a specified area by stabilizing the tip-surface distance in each point $(x, y)$ at voltage $V_{stab}$ and current $I_{stab}$, switching off the feedback and recording a $dI/dV$ curve from $V_{start}$ to $V_{end}$ ($V_{start} \leq V_{stab}$) (ref 6). As a result, WF mapping produces a three-dimensional array of $dI/dV$ data, which allows (i) obtaining spatially resolved $dI/dV$ images at different values of $V_{sample}$ and (ii) extracting the $dI/dV$ spectra at specific positions corresponding to specific topographic features.

The sample structure and experimental setup are sketched in Figure 1a. Large-scale constant-current STM images (not shown) revealed a QD density of $2.5 \times 10^{10}$ cm$^{-2}$, while the shape of the QDs was determined from STM images acquired within a smaller area, as in Figure 1a. Despite some variation in size (with a typical lateral extension of 30 nm along both $[110]$ and $[1\bar{1}0]$ directions and an average height of 5-6 nm), all QDs exhibit similar facets. As shown in Figure 1a, the QDs have a pyramidal shape with well-defined facets and a fairly sharp summit. We observed a pronounced shape anisotropy as visible from the three-dimensional (3D) view of Figure 1a as well as from the height profiles reported in Figure 1b. In particular, the height profile is triangular along $[110]$, and rounded along the perpendicular direction $[1\bar{1}0]$. The inclination angle between the facets and the substrate is approximately 19°, in line with (114) planes. Similar structural features and anisotropic shapes were reported in recent studies, in particular by Marquez *et al.*[18]. Besides the bright QDs, several steps are visible on the wetting layer (WL), which has a $2 \times 4$ reconstructed structure similar to that of the GaAs(001) surface[6].

To examine the electronic structure in our QDs, scanning tunneling spectroscopy was performed on individual dots. Figure 2 (top part) shows typical $(dI/dV)/(I/V)$ spectra taken on a relatively small QD in different positions. Four clearly resolved peaks corresponding to resonances in the QD's spectral density were observed in the STS spectra, indicating that the effective energy resolution $\delta E$ is smaller than the level spacing. In particular, we identify sharp peaks marked A at 840 mV, B at 1040 mV, C at 1140 mV, and a broader peak D at 1370 mV, whose full widths at half maximum (FWHM) are about 30, 25, 40, and 75 mV, respectively. Moving from the QD center to its sides, the intensity of the low-energy peak decreases while the others increase in weight. Moreover, we observe a small blue shift of the whole spectrum to higher energies, probably due to the increased band bending at a smaller distance between the tip and the degenerately doped GaAs backgate when the tip is at the rim of the QD[6]. However, since the resulting peak shifts are small, $dI/dV$ images still largely represent the peak intensity as a function of position[6]. For comparison, the STS curves on the wetting layer (outside the QD, cyan curve of Figure 2) are featureless in the same voltage range. Thus, we ascribe the $(dI/dV)/(I/V)$ peaks to quantized states of the QD.

The spatially resolved mapping allows us to determine the symmetry of the corresponding squared WFs by revealing their shape. The bottom part of Figure 2 shows the 3D spatial variation of the $(dI/dV)/(I/V)$ signal at tunneling voltages where peaks in the QD's spectral density are exhibited. In detail, we observe the following approximate symmetries: one *s*-like (A), two (or possibly three) *p*-like (B, C, and D), on going from low to high energy. As expected for the moduli squared of the two lowest-energy one-electron *s*- and *p*-like orbitals, states A and B exhibit a roughly circular symmetric intensity distribution and elongation along the $[1\bar{1}0]$ direction with a node in the center, respectively. Unexpectedly, state C shows again a *p*-like symmetry in the $[1\bar{1}0]$ direction, as before, instead of $[110]$ as expected for the second *p*-like orbital[6]. As a consequence, it is not possible to explain the WF sequence (and map C in particular) in terms of states of the systems with just one electron, since in this case we would expect the appearance of either a single *p*-state or two *p*-states elongated in the $[1\bar{1}0]$ and [110] directions, respectively. On the other hand, we also exclude the charging of the same *p*-like orbital with a second electron having opposite spin, since in this case we would expect to observe also the charging of the *s*-like orbital, resulting in a second state with circular symmetry. For similar reasons, C is not a phonon replica[22]. As a consequence, we believe that correlation effects play a key role and the quasi particle WF concept is needed to understand these features. We do not observe any *p*-state elongated along [110] in the voltage range explored since it lies at high energy due to the strong anisotropy of our QDs (see Figure 1), according to previous theoretical predictions[23].

For further discussion of the state sequence, we start by noting that states B, C, and D are oriented in the same direction ($[1\bar{1}0]$), as we also observed in our previous report[6]. This indicates that the effective potential confining the electrons has a $C_{2v}$ symmetry, i.e., is approximately an elliptic harmonic trap in the $(x, y)$ plane where the elliptical anisotropy models both geometrical deviations from perfect circularity and atomistic effects due to strain, piezoelectric fields, and interface matching. FCI calculations[17] taking into account this QD anisotropy, the effect of dielectric environment, and electron correlation were thus performed to address the effect of electron-electron interaction on quasi particle WF mapping. In more detail, we considered the fully interacting Hamiltonian for different electron numbers, *N*, and we solved the corresponding few-body problems numerically in an accurate way, by means of the FCI method[17], that we previously have successfully applied in predicting QD transport[24] and Raman spectra[7]. Eventually, from the FCI output for *N* and *N – 1* electrons, consisting in correlated ground and excited states expanded into a superposition of different electronic configurations (Slater determinants), we compute the quasi particle WF maps[9, 10] to be compared with states of Figure 2 (ref 25). The more correlated the dot, the stronger the configuration mixing, which modifies the spatial shape of the quasi particle WF due to interference among different orbitals. In this respect, it is worth noting that the effect of dielectric mismatch between InAs and vacuum is believed to be strong[22, 26, 27]. Qualitatively, its inclusion into our model implies that[26, 27] (i) the single-particle confinement potential is changed (self-polarization effect) and (ii) the electron-electron interaction is strengthened (interaction with surface image charges of like sign). While issue (i) does not change qualitatively the model, feature (ii) enforces correlation effects, which are relevant in the present context[22]. In order to qualitatively mimic this effect in the calculation, we thus introduced an *ad hoc* strengthening factor multiplying the Coulomb interaction term of the Hamiltonian. Specifically, the equivalent relative dielectric constant $\kappa$ used corresponds to $\kappa = 4$. While a rigorous treatment of effects (i) and (ii) is possible (see ref 20), it is beyond the aim of the present work[22].

In Figure 3a the calculated STS map for the ground state → ground state tunneling transition $N=1 \rightarrow N=2$ (being $N$ the number of electrons in the QD before and after the tunneling transition) is shown in the correlated case for increasing values of the dot anisotropy (from left to right). For a perfectly circular QD (left plot) we unambiguously predict an isotropic $s$ symmetry for the WF (in the non-interacting case the STS map is only slightly squeezed with respect to the correlated image on the left-hand side of Figure 3a). On the other hand, by increasing the QD eccentricity, we see that progressively the interacting WF forms two peaks along the major axis, while its noninteracting counterpart is simply a Gaussian being elongated along the same direction (not shown). We thus identify the state C in WF mapping with this tunneling transition $N=1 \rightarrow N=2$ from ground to ground state. The formation of the two peaks, together with the loss of weight at the dot center, is due to the destructive interference between $s$ and $d$ states of the harmonic oscillator along the major axis (belonging to the same representation $A_1$ of the $C_{2v}$ group), which is a correlation effect. Such an effect is ultimately connected with the loss of angular correlation induced by the breaking of circular symmetry, which is compensated by radial correlation along the major axis. This can also be seen as a manifestation of the general statement that the importance of correlation increases as the system dimensionality is reduced (in this case, from 2D to 1D).

In Figure 3b, we present (from left to right) the experimental STS energy spectrum (first column) and typical predicted maps calculated separately for the charging processes corresponding to the injection of the first (second column) and second (third column) electron into the QD. Different FCI calculations for different sets of input parameters were performed, showing the same qualitative trend as displayed in Figure 3. Note that the tip-induced band bending has to be taken into account to transform the voltage scale into the energy scale. Specifically, the comparison of experimental energy spacings (left diagram) with those calculated (center and right) suggests that states A and B can be ascribed to the two lowest-energy states predicted for the tunneling process $N=0 \rightarrow N=1$, while states C and D could be associated to the tunneling process $N=1 \rightarrow N=2$. Moreover, the tunneling regime appears to switch from $N=0 \rightarrow N=1$ to $N=1 \rightarrow N=2$ at larger voltages/energies due to the increased current flowing in the tunneling junction.

Thus, in this case the quasi-particle WFs probed by $dI/dV$ mapping considerably deviate from the noninteracting WFs. As far as the energy scale is concerned, theoretical results are consistent with experimental ones if a voltage/energy conversion factor around 2.6 is assumed, while a factor around 2 is obtained from a lever-arm-rule estimation considering that the applied voltage drop is shared between a 5-6 nm high QD and a 5 nm tunnel barrier while the tip-sample distance is typically around 1 nm. However, a strict quantitative comparison between measured and predicted quantities is presently out of reach because sample details, such as the exact confinement potential, and the degree of anisotropy are unknown. Since the peak FWHM in the WF mapping is around 50 mV, the energy resolution is not sufficient to distinguish the calculated ground- $\rightarrow$ (ground-) excited-state transitions (right column of Figure 3b) $\alpha$ from $\beta$ (experimental state C) and $\gamma$ from $\delta$ (D) (ref 22). A comparison of the experimental line profiles (Figure 3c left, showing the maps of states B, C, and D along a QD volume slice) and theoretical ones (Figure 3c right, for state $p_x$, and overlaps of α and β, γ and δ, respectively) reinforces this conclusion. Nevertheless, the excellent agreement between the measured and predicted solid blue curves of Figure 3c demonstrates that the spatial modulation of map C is a genuine effect of Coulomb interaction: In fact, the overlap (dashed blue line) of the noninteracting counterparts of states α and β -the s and $p_x$ orbitals, mixed together with the same 1:1 ratio as α and β- poorly compares to C.

Our interpretation is further supported by a study of the dependence of the STS spectra on the stabilization current. Since the tunneling rate through the undoped GaAs tunnel barrier is much larger than the typical tunneling rate from the tip to the QD[6], in order to systematically investigate charging effects we recorded tunneling spectra at different stabilization currents $I_{stab}$ (i.e., we varied the tip-dot distance) to gradually increase the tunneling rate into the QD in order to populate it[28]. In Figure 4, we focus on states B and C, which are visible at voltages between 1040 and 1140 mV, dominating the spectra acquired on the QD sides. In the $dI/dV$ spectra collected at increasing $I_{stab}$, the first (second) peak corresponding to the first (second) *p*-like orbital gradually disappears (appears). This suggests that at low (high) values of the stabilization current we probe the energy spectrum of an uncharged (charged) quantum dot, while for intermediate values we can see both peaks (as in our WF maps). Thus, the state C must be associated with the tunneling process $N=1 \rightarrow N=2$ since it appears at higher $I_{stab}$ (the STS levels measured at increasing values of the current correspond to higher values of QD occupancy). On the other hand, a discussion of state D is more complicated. It could be tentatively ascribed to the tunneling process $N=1 \rightarrow N=2$ (states $\gamma$ and $\delta$ in Figure 3b) according to theoretical results. However, from Figure 2 peak D appears to be located at an energy where the wetting layer significantly contributes to the spectral density (which is high also outside the QD). So, a coupling is possible and this could explain the increased FWHM observed for state D (Figure 2).

In conclusion, we have investigated the role of electron-electron interaction in few-particle QDs by WF mapping. Correlation effects were found to be relevant for tunneling spectroscopy at high tunneling currents, and a many-body picture was developed to explain these features on the basis of theoretical calculations. These results demonstrate the sensitivity of STS to electron correlation. They could inspire experiments for a broad range of nano-objects including carbon nanotubes and single molecules.

**Acknowledgments:**

The authors gratefully acknowledge financial support by the EU-network project "Nanospectra", the Deutsche Forschungsgemeinschaft (SFB508, TP A6 and B7), the INFM-CINECA Supercomputing Project 2006, and the MIUR-FIRB Project RBIN04EY74. Moreover, M.R. would like to thank S. Corni and C. Goldoni for fruitful discussions.

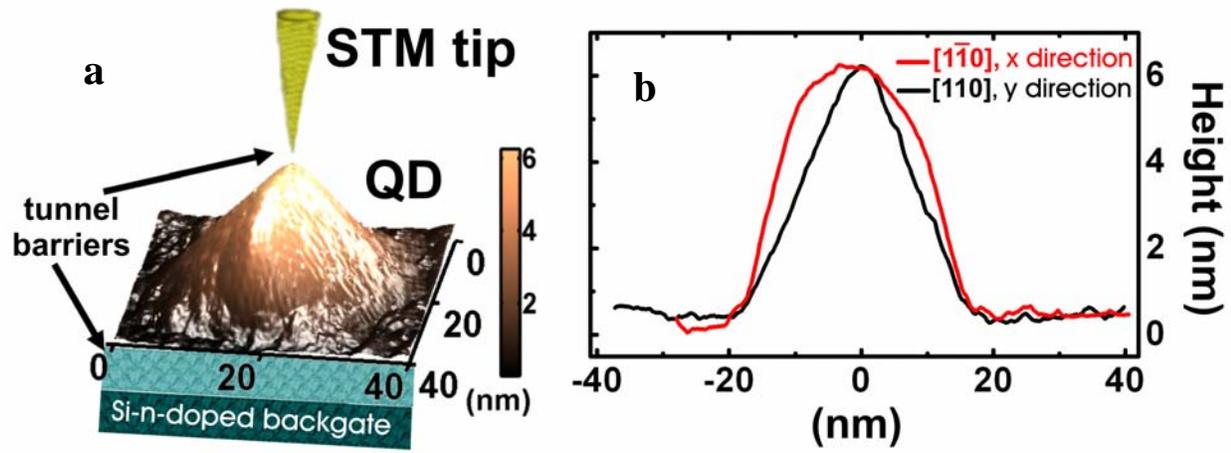

**Figure 1 Experimental setup and dot morphology.** (**a**) Experimental setup and three-dimensional STM image of a representative uncapped QD grown on n-doped GaAs(001) substrate. (**b**) Height profiles of the same QD along two perpendicular axes, parallel to $[110]$ and $[1\bar{1}0]$ directions, respectively.

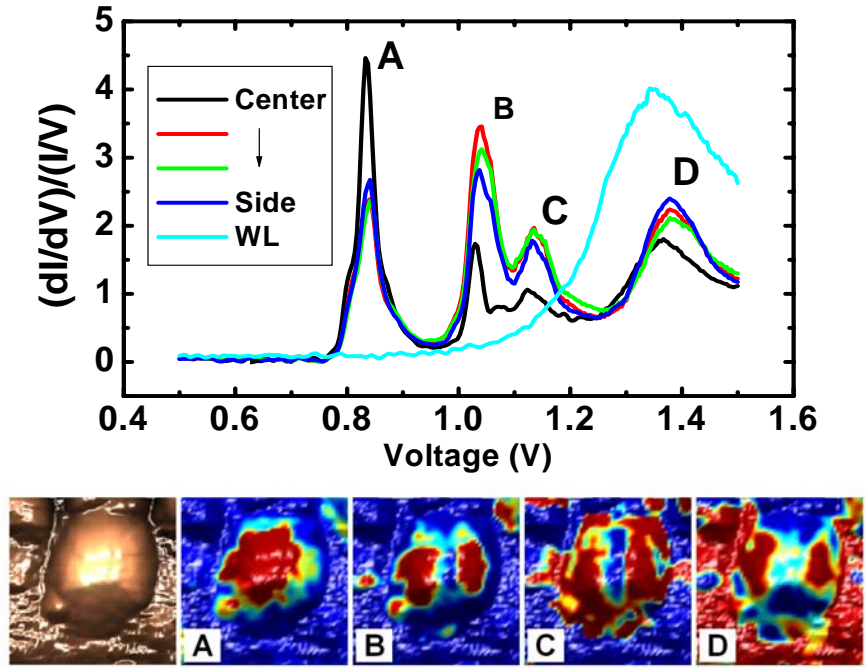

**Figure 2 STS spectra and wave function maps.** Top panel: $(dI/dV)/(I/V)$ spectra vs $V$ measured at different positions on a single QD, moving from the QD center to its sides. Bottom panel: STS spatial maps of a single representative dot, taken at 840, 1040, 1140, and 1370 mV, for resonances A, B, C, and D, respectively (second-fifth panel). The color code represents the STS signal with respect to the topographic STM image on the left-hand side (first panel), increasing from blue to red. The dot height in the first panel varies from 0 (dark brown) to $\approx 6$ nm (light brown). The lateral extension of all maps is $30 \times 30$ nm. $I_{stab} = 100$ pA, $V_{stab} = 1.5$ V, $V_{mod} = 4$ mV.

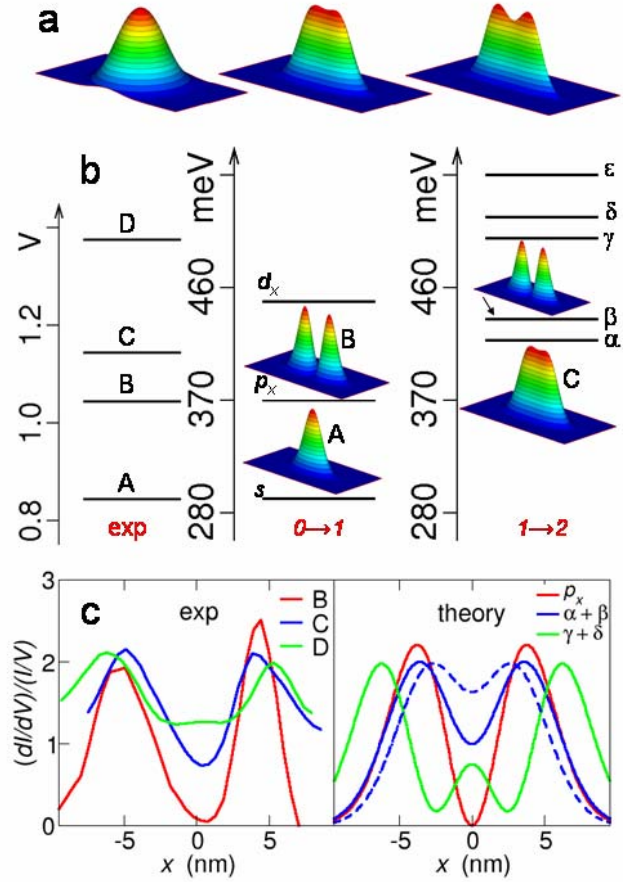

**Figure 3 Comparison between measured and predicted wave function maps.** (**a**) Calculated STS maps (WF square moduli) for the ground state → ground state tunneling transition $N=1 \to N=2$, as a function of dot anisotropy (the central image is labeled C in panel b). From left to right, the ratios of the lateral extension of the major to minor axes are 1, 2.5, and 5, respectively. (**b**) Experimental STS energy spectrum (left column) and calculated states for the tunneling processes $N=0 \to N=1$ (center column) and $N=1 \to N=2$ (right column). The predicted images of experimentally relevant states are also shown. The in-plane size of all 3D plots is $4 \times 2$ units of the lateral extension of the noninteracting *s* orbital in the elongated direction. (**c**) Profiles of STS maps (left) and predicted probability densities (right, in arbitrary units) along a QD volume slice. The measured profiles were extracted from the data of Figure 2 after averaging in the transverse direction (see Figure 2 in Supporting Information). The predicted solid blue (green) curve corresponds to the overlaps of α and β (γ and δ) states, mixed with a 1:1 ratio. The dashed blue line is the 1:1 overlap of *s* and $p_x$ noninteracting orbitals.

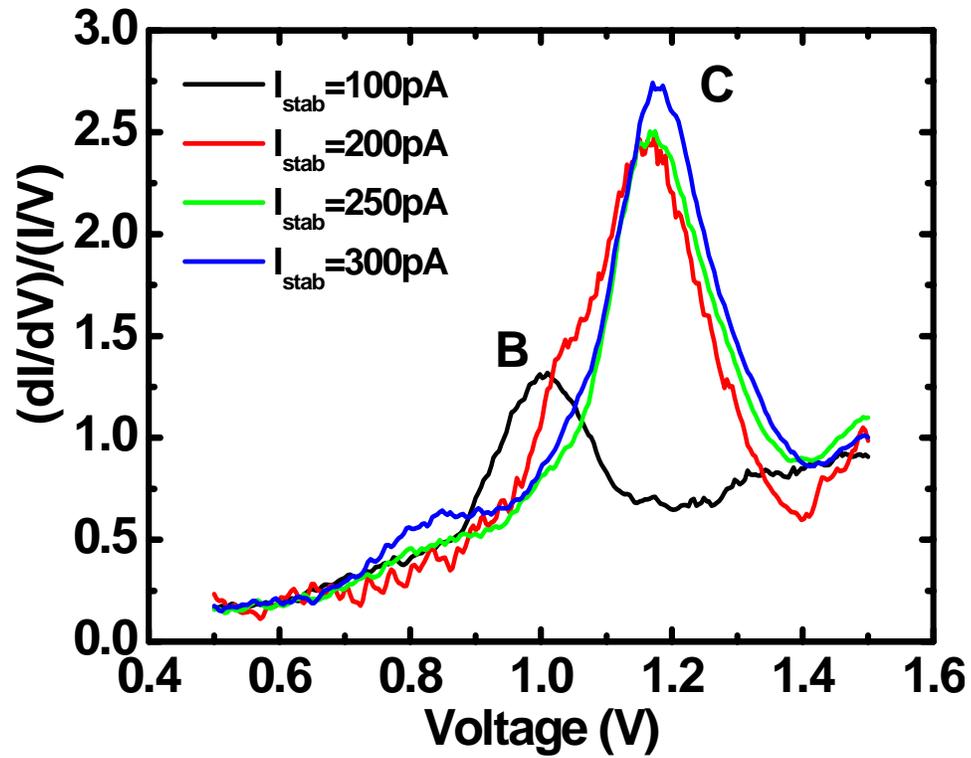

**Figure 4  Changing the dot occupancy by varying the stabilization current.** $(dI/dV)/(I/V)$ tunneling spectra vs $V$ as a function of the stabilization current $I_{stab}$ measured at the QD side, where states B and C are dominant.

# Supplementary Discussion and Figures

**1. Change of local density along the QD**

For further comparison of experimental and theoretical data, we provide in Supplementary Figure 1 the separate linear profiles of calculated states α, β, γ, and δ. The 1:1 mixing of pairs α and β, γ and δ, respectively, is shown on the right hand side of Fig. 3c. The experimental linear profiles appearing on the left hand side of Fig. 3c were obtained by averaging the differential conductance in the transverse direction in the spatial regions shown in Supplementary Figure 2.

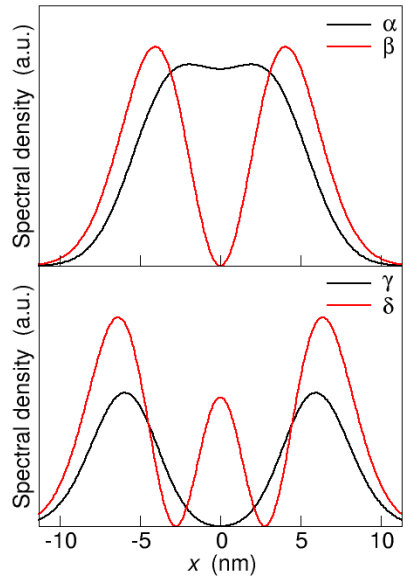

**Supplementary Figure 1.** Separate profiles extracted from α and β (top) and γ and δ (bottom) predicted maps of Fig. 3b.

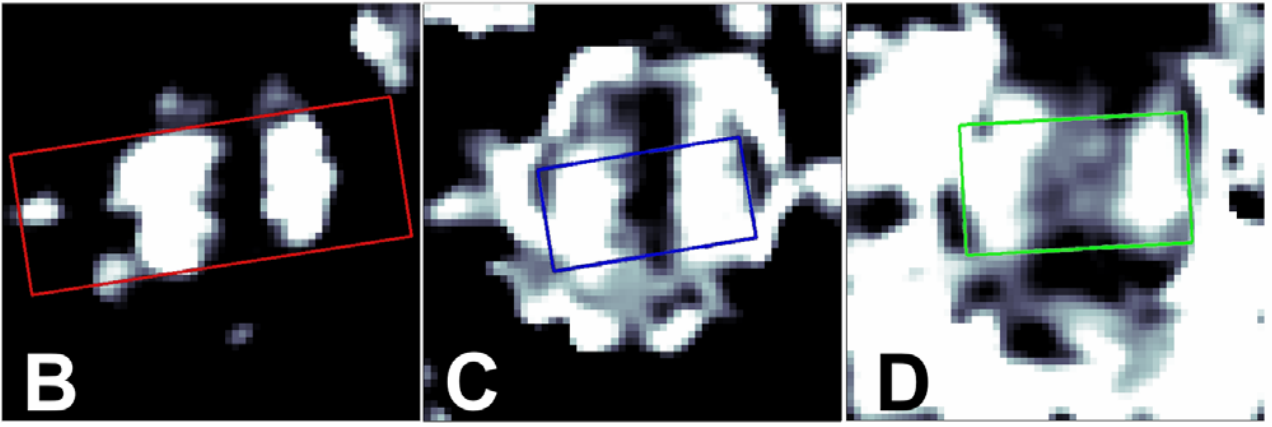

**Supplementary Figure 2.** STS spatial maps (same set of data as the maps of Fig. 2) of a single representative dot recorded at the voltages 1040 (B), 1140 (C), and 1370 (D) mV, respectively. The size of all maps is 30 × 30 nm. $I_{stab}$ = 100 pA, $V_{stab}$ = 1500 mV, $V_{mod}$ = 4 mV. White (black) colour stands for high (low) values of $(dI/dV)/(I/V)$. The colored rectangles show the spatial regions where the profiles in Fig. 3c were averaged.

## 2. Interpretation of state C

We associate states C and D to the tunneling process $N=1 \rightarrow N=2$. In particular, since the peak FWHM in the WF mapping is around 50 mV, the energy resolution is not sufficient to distinguish the calculated ground- → (ground-) excited-state transitions (right column of Fig. 3b) $\alpha$ from $\beta$ (experimental state C) and $\gamma$ from $\delta$ (D). By assuming a value 2.6 for the voltage/energy conversion factor, the B to C energy splitting is about 38 meV, quite close to that of optical phonons in the InAs/GaAs system. However, we exclude that state C is a phonon replica of B (involving the emission of an optical phonon of the QD or of the GaAs layer below) for the following three reasons.

**(i)** The dependence of the STS spectra on the stabilization current suggests that state C must be associated to the tunneling process $N=1 \rightarrow N=2$ since it only appears at higher $I_{stab}$ where state B then disappears (the STS levels measured at increasing values of the current correspond to higher values of QD occupancy). On the other hand, we would not expect state B to disappear if it were the main peak associated to elastic tunneling.

**(ii)** A phonon replica would presumably appear as a small side peak due to a reduction of transition probability (it is a second order process), while B and C states exhibit comparable spectral densities (see Fig. 2).

**(iii)** We did not observe any phonon replica of state *s*.

Moreover, the small energy splitting between B and C states along with the small blue shift of the whole spectra to higher energies on the QD sides accounts for the high intensity regions on the QD sides visible in the STS spatial map of state C (Figure 2b and Supplementary Figure 2), which belong to state B.

### 3. Effect of the dielectric mismatch among quantum dot, vacuum, and STS tip

As mentioned in the main text, the dielectric mismatch between QD and vacuum is responsible for the appearance of negative image charges on the dot surface, which increases the effective Coulomb repulsion among electrons in the QD. This effect is described in a simple way by the effective value $\kappa = 4$ of the relative dielectric constant used in the FCI calculation: the smaller $\kappa$, the larger the Coulomb matrix elements, the stronger the electron correlation. A concern regards the presence of the STS tip, which in principle could screen the interaction and wipe off the effect. In this section we discuss the competing effects of the QD-vacuum and vacuum-electrode interfaces, and conclude that the Coulomb repulsion is significantly enhanced even in the presence of the STS tip.

The estimate of the effective Coulomb repulsion relies on the calculation of the Green's function as defined in classical electrostatics [see e.g. Jackson, J. D. *Classical Electrodynamics*, Chapters 2 and 3 (Wiley, New York, 1975), and Refs. 26, 27]. Since only the simplest geometries allow for an almost analytical solution of the problem, here we consider a highly schematic model of the dielectric environment consisting in three parallel layers indefinitely extended in the lateral directions, namely two dielectrics (with relative dielectric constants $\kappa_1$ and $\kappa_2$, respectively) and a metal. Let $z$ be the axis perpendicular to the perfectly flat interfaces, and locate the dielectric 1 / dielectric 2 interface at $z = 0$ and the dielectric 2 / metal interface at $z = L$ [the limit $z \to -\infty$ $(+\infty)$ corresponds to the bulk dielectric 1 (metal)]. The calculation proceeds through the lines indicated e.g. in Panofsky, W. K. H. & Phillips, M. *Classical Electricity and Magnetism,* Sec. 5.8 (Addison-Wesley, New York, 1962), based on the matching of the solutions of the Laplace equation in the three media by exploiting the lateral translational symmetry. Here we only state the final result for the effective Coulomb repulsion $V_{\text{eff}}(\rho, z)$ between two electrons, located at relative lateral distance $\rho$ and both placed at $z < 0$:

$$V_{\text{eff}}(\rho, z) = \frac{e^2}{\kappa_1} \int_0^\infty dk [1 + f_z(k)] J_0(k\rho),$$

$$f_z(k) = \frac{(\kappa_2 - \kappa_1)\exp[2k(z+L)] + (\kappa_2 + \kappa_1)\exp(2kz)}{\kappa_1[1 - \exp(2kL)] - \kappa_2[1 + \exp(2kL)]},$$

where $J_0(x)$ is the zero-order Bessel function of the first kind. Note the two important cases: (i) With $\kappa_1 = \kappa_2$, $L \to \infty$, we have $f_z(k) \to 0$ and recover the bulk dielectric result $V_{eff}(\rho, z) \to e^2/(\kappa_1 \rho)$. (ii) With $L \to \infty$ we recover the result for the two-dielectric problem, $V_{eff}(\rho, z) \to e^2/(\kappa_1 \rho) + e^2(\kappa_1 - \kappa_2)/[\kappa_1(\kappa_1 + \kappa_2)(\rho^2 + 4z^2)^{1/2}]$, given by the sum of the bulk pair interaction plus the repulsion with the image charge placed at vertical distance $|2z|$. The maximum value that $V_{eff}$ may reach is obtained in the latter case by putting $\kappa_1/\kappa_2 \to \infty$ and $z = 0$, resulting in twice the bulk value (such an upper bound is specific to the planar plate geometry).

To proceed further, we assume that electrons are laterally confined by a circular harmonic potential $V_{conf}(\rho) = m\omega^2 \rho^2/2$, where $\hbar\omega$ is the confinement energy and $m$ is the effective electron mass. The (Fock-Darwin) ground state wave function of the single-particle problem, neglecting the motion along $z$, is the $s$-like Gaussian $\varphi_s(\rho) = \exp(-\rho^2/2\ell_{QD}^2)/\sqrt{\ell_{QD}^2 \pi}$, where $\ell_{QD} = \sqrt{\hbar/m\omega}$ is the characteristic lateral extension of the orbital. In order to quantify the strength of electron correlation, we compute the Hubbard-like electrostatic repulsion $U$ referred to the $s$-orbital:

$$U = \iint d\vec{\rho}' d\vec{\rho} |\varphi_s(\rho')|^2 V_{eff}(|\vec{\rho}' - \vec{\rho}|, z) |\varphi_s(\rho)|^2.$$

After separation of the relative and center-of-mass coordinates, we eventually obtain an expression convenient to numerical evaluation:

$$U = \frac{e^2}{\kappa_1 \ell_{QD}} \int_0^\infty dx [1 + f_z(x/\ell_{QD})] \exp(-x^2/2).$$

To apply the latter equation to the actual experimental geometry, we introduce effective thicknesses for the QD and vacuum regions, respectively. We proceed as follows: We model the QD as half an ellipsoid with basal semi-axes $a = b = 15$ nm, vertical semi-axis $c = 5$ nm, and define the equivalent QD height $h_{equiv}$ as the height of the cylinder of radius $a$ and same volume as that of the ellipsoid: $h_{equiv} = 2c/3$. We then assume that electron motion is confined in the middle of the QD, $z = -h_{equiv}/2$, and put $L = 1$ nm. The value of $U$ obtained from the above parameters has to be compared with the reference case in which the STS tip is absent, i.e. $L \to \infty$. The relative reduction turns out to be 29% (27%) for GaAs (InAs). Nevertheless, $U$ is still 5% (9%) larger than the value corresponding to a unique bulk, i.e. $\kappa_1 = \kappa_2$, $L \to \infty$. This demonstrates that the enhancement of electron correlation due to the dielectric mismatch survives to the screening effect of the STS tip. Besides, correlation effects are expected to be much larger than those estimated within this simple model. In fact, while in the planar plate geometry the maximum value of $U$ attainable is only twice the bulk value, in the case of a spherical dot in the vacuum $U$ may be several times larger than the bulk value without any interface [Franceschetti, A., Williamson,

A. & Zunger, A. Addition spectra of quantum dots: the role of dielectric mismatch. *J. Phys. Chem. B* **104,** 3398-3401 (2000)]. The representation of the actual experimental setup should fall somewhere in the middle between the limiting cases of the planar plate and spherical geometries, respectively. In addition, in our calculation the effective dot-tip distance is likely underestimated.

In conclusion, on the basis of a rough estimate we believe the actual value of $\kappa$ employed in the FCI calculation is sensible.